\documentclass[a4paper,fleqn,usenatbib,useAMS]{mnras}
\usepackage{graphicx}

\graphicspath{ {ImageI/} }
\usepackage{amsmath}	
\usepackage{amssymb}	
\usepackage{multicol}   
\usepackage{bm}		    
\usepackage{pdflscape}	

\usepackage[T1]{fontenc}
\usepackage{ae,aecompl}
\usepackage{mathptmx} 

\title[A \textit{Kepler} Eclipsing Binary with Magnetic Activity and Hybrid Pulsations. I. Binary modelling]
{KIC 6048106: An Algol-type Eclipsing System with Long-term Magnetic Activity and Hybrid Pulsations. I. Binary modelling}
\author[Samadi Gh., et~al.]{
A. Samadi Gh.,$^{1,2}$\thanks{Visiting Researcher at ROB; E-mail: samadi.aunia@tabrizu.ac.ir; samadi.aunia@gmail.com;}
P. Lampens,$^{2}$ 
D. M. Jassur,$^{1}$
\\
$^{1}$ Department of theoretical Physics and Astrophysics, Physics Faculty, University of Tabriz, P.O.Box 51664, Tabriz, Iran\\ 
$^{2}$ Koninklijke Sterrenwacht van Belgi\"e, Ringlaan 3, B-1180 Brussel, Belgium\\}
\date{Last updated 2015 May 22; in original form 2013 September 5}

\pubyear{2017}

\begin{document}
\label{firstpage}
\pagerange{\pageref{firstpage}--\pageref{lastpage}}
\maketitle

\begin{abstract}
The A/F-type stars and pulsators ($\delta$ Scuti-$\gamma$ Dor) are in a critical regime where they experience the transition from radiative to convective transport of energy in their envelopes. Such stars can pulsate in both gravity and acoustic modes. Hence, the knowledge of their fundamental parameters along with their observed pulsation characteristics can help in improving the stellar models. When residing in a binary system, these pulsators provide more accurate and less model dependent stellar parameters than in the case of their single counterparts. We present a light curve model for the eclipsing system KIC 6048106 based on the \textit{Kepler} photometry and the code PHOEBE. We aim to obtain accurate physical parameters and tough constraints for the stellar modelling of this  intermediate-mass hybrid pulsator. We performed a separate modelling of three light curve segments that show a distinct behaviour due to a difference in activity. We also analysed the \textit{Kepler} ETVs. KIC 6048106 is an Algol-type binary with F5-K5 components, a near-circular orbit and a 1.56-d period undergoing variations of the order of $\frac{\Delta P}{P}\simeq 3.60\times10^{-7}$ in $287\pm7$ days. The primary component is a main-sequence star with $M_\mathrm{1}=1.55\pm0.11M_{\sun},~R_\mathrm{1}=1.57\pm0.12R_{\sun}$. The secondary is a much cooler subgiant with $M_\mathrm{2}=0.33\pm0.07M_{\sun},~R_\mathrm{2}=1.77 \pm0.16R_{\sun}$. Many small near-polar spots are active on its surface. The second quadrature phase shows a brightness modulation on a time scale $290\pm7$ days, in good agreement with the ETV modulation. This study reveals a stable binary configuration along with clear evidence of a long-term activity of the secondary star.
\end{abstract}
\newpage
\section{Introduction}\label{sect:intro}
The intermediate-mass A-F type stars of mass $ 1.5$-$2.5M_{\sun} $ play an important role in the study of stellar structure and evolution. An important issue (which needs better understanding) is the fact that they are passing through a transition phase: the two inefficient thin convective zones in their outer layers of the H and HeII partial ionization zones merge to form a larger efficient convective zone \citep{Dupret2007}. There are always some degeneracies in determining free parameters through stellar modelling for stars in such transition phases. The confrontation of the observations with theoretical models is needed to increase the knowledge of some stellar parameters. Furthermore, chemical composition and mixing processes along with other fundamental parameters will be better determined. In the H-R diagram, these stars lie inside the classical Cepheid instability strip and its intersection with the main sequence. The study of pulsations in that zone unveils much about mass, age, chemical structure, mixing processes, internal rotation profiles, and the size of convective zones along with the mechanisms of the mode excitation and so on. The characterization of stellar pulsations in the above mentioned region is a great tool to study their internal chemical structure and rotation profile in order to find the missing pieces of stellar modelling puzzles.\\
$ \delta $ Scuti stars are main-sequence stars with masses in the range 1.5-2.5$M_{\sun}$ and F5-A2 spectral types and effective temperatures of $ T_\mathrm{eff} $ = 6700-8900 K. Their logarithmic luminosity range is $ \log{(L/L_{\sun})}$ = [0.6,2.0]. $ \delta $ Scuti stars are population I stars (they have population II counterparts which are called SX Phoenicis stars) located at the extension of the Cepheid instability strip toward the lower luminosities and overlapping with the main sequence stage undergoing hydrogen core or shell burning.\\
$\gamma$ Doradus stars are slightly cooler and less massive than $ \delta $ Scuti stars. These stars with F5-A7 spectral types, $ T_\mathrm{eff}$ = 6700-7400 K, and masses around 1.5-1.8$~M_{\sun} $ are located a bit above the main sequence and cross the red edge of the classical instability strip, between $ \delta $ Scuti instability strip and solar-like pulsators. For a better view of the location of the $\gamma$ Doradus and $\delta $ Scuti instability strips and their overlap we refer to Figure 2 in~\citet{Dupret2005}. They have luminosities ranging from $ \log{(L/L_{\sun})}$= [0.7,1.1]. The F-type stars have projected rotational velocities around 50 $~\mathrm{kms}^{-1} $.\\
A new issue has risen with the \textit{Kepler} discovery of these hybrid pulsators all over the $\delta $ Scuti instability strip since such hybrid stars are not predicted at the hot edge by theory~\citep{Grigahcene2010}. Other excitation mechanisms besides $ \kappa $ and convective flux blocking might be needed in order to explain the variety of pulsation modes in hybrid stars. This allows us to improve upon the stellar structure and evolutionary models for intermediate-mass stars which are in the transition phase. In addition simultaneous core-surface information can be obtained from the hybrid $ \delta $ Scuti-$ \gamma $ Dor stars. Thus, the study of hybrids can provide information about the structure from core to surface layers as well as reveal the entire profile of the internal rotation.\\
Furthermore, when hybrid pulsators reside in an eclipsing binary system, the system offers multiple additional constraints for studies of stellar structure and evolution. In this case, we can use the advantages of a system with two components having the same age and chemical composition. Whenever an eclipsing binary shows $ \delta $ Scuti-$ \gamma $ Doradus pulsation in one or both components, we may acquire a collection of accurately determined fundamental, atmospheric and orbital parameters. The determination of these parameters is based on the proper adjustment of the light and radial velocity curves. Here the model-independent stellar parameters which are derived from an observational study of $\delta $ Scuti-$ \gamma $ Doradus components in eclipsing binaries would improve the reliability of the models. Statistical studies on large samples are needed to fully resolve global and internal stellar  properties for the mentioned group.\\
It is the bulk of space photometric data with unprecedented accuracy which is paving the way of aforementioned researches. The light curves from \textit{Kepler} with an accuracy in the order of $\mu$ mag can reveal the very fine variations in flux as a finger-print of $ \delta $ Scuti-$ \gamma $ Dor pulsations. Detection of the low-frequency g-modes and high frequency p-modes with amplitudes lower than 0.1~mmag is made possible thanks to highly accurate continuous data sets from space missions such as \textit{MOST}~\citep{Walker2003}, \textit{CoRoT}~\citep{Auvergne2009}, and \textit{Kepler}~\citep{Koch2010}.\\
The studies on oscillating eclipsing binaries goes back to \citet{Mkrtichian2002} and \citet{Mkrtichian2003}. Other similar studies, on Algol-type binaries, were done by \citet{Soydugan2006} detecting 25 Algol systems with confirmed $ \delta $ Scuti pulsations. \citet{Christiansen2007} announced the first high amplitude $ \delta $ Scuti pulsations in the eclipsing binary system UNSW-V-500. The studies done on large samples of \textit{Kepler} and \textit{CoRoT} revealed, for the first time, that hybrid behaviour is common among the A-F type pulsators,~\citep{Grigahcene2010} and~\citep{Hareter2010}. Several studies concern individual stars in eclipsing binary systems with hybrid $ \delta $ Scuti and $ \gamma $ Dor pulsations e.g. 'KIC 10661783', \citet{Southworth2011}, 'CoRoT 102918586', \citet{Maceroni2013}, 'KIC 11285625', \citet{Debosscher2013}, 'KIC 4844587', \citet{Hambleton2013}, 'DY Aqr', \citet{Alfonso-Garzon2014}, 'KIC 3858884', \citet{Maceroni2014}, 'CoRoT 105906206', \citet{da Silva2014}, 'KIC 8569819', \citet{Kurtz2015a}, 'KIC 10080943' \citet{Schmid2015} and 'KIC 9851944' \citet{Guo2016}. It is obvious that we are still far from obtaining statistically relevant conclusions on the hybrids due to the low number of targets studied.\\
This study focuses on KIC 6048106, which is a semi-detached, Algol-type eclipsing binary star with $P_\mathrm{orb}$=1.56 d. It shows pulsations of type $ \gamma$ Doradus-$ \delta $ Scuti. A preview of the \textit{Kepler} photometric data along with its characteristics is presented in Section~\ref{sec:obs}. The methodology of the research both for the binary modelling (paper I) and the pulsation study (paper II) is explained in Section~\ref{sec:method}. In Section~\ref{sec:res}, we present the overall results of the binary modelling and of the ETV curve study in detail. The overall conclusion from the Binary modelling is discussed in Section ~\ref{sec:con}. The results and the discussion on the pulsational behaviour is presented in paper II. Pulsation study.
\section{Observations and \textit{Kepler} Photometry}
\label{sec:obs}
\begin{figure}
	\includegraphics[width=\columnwidth]{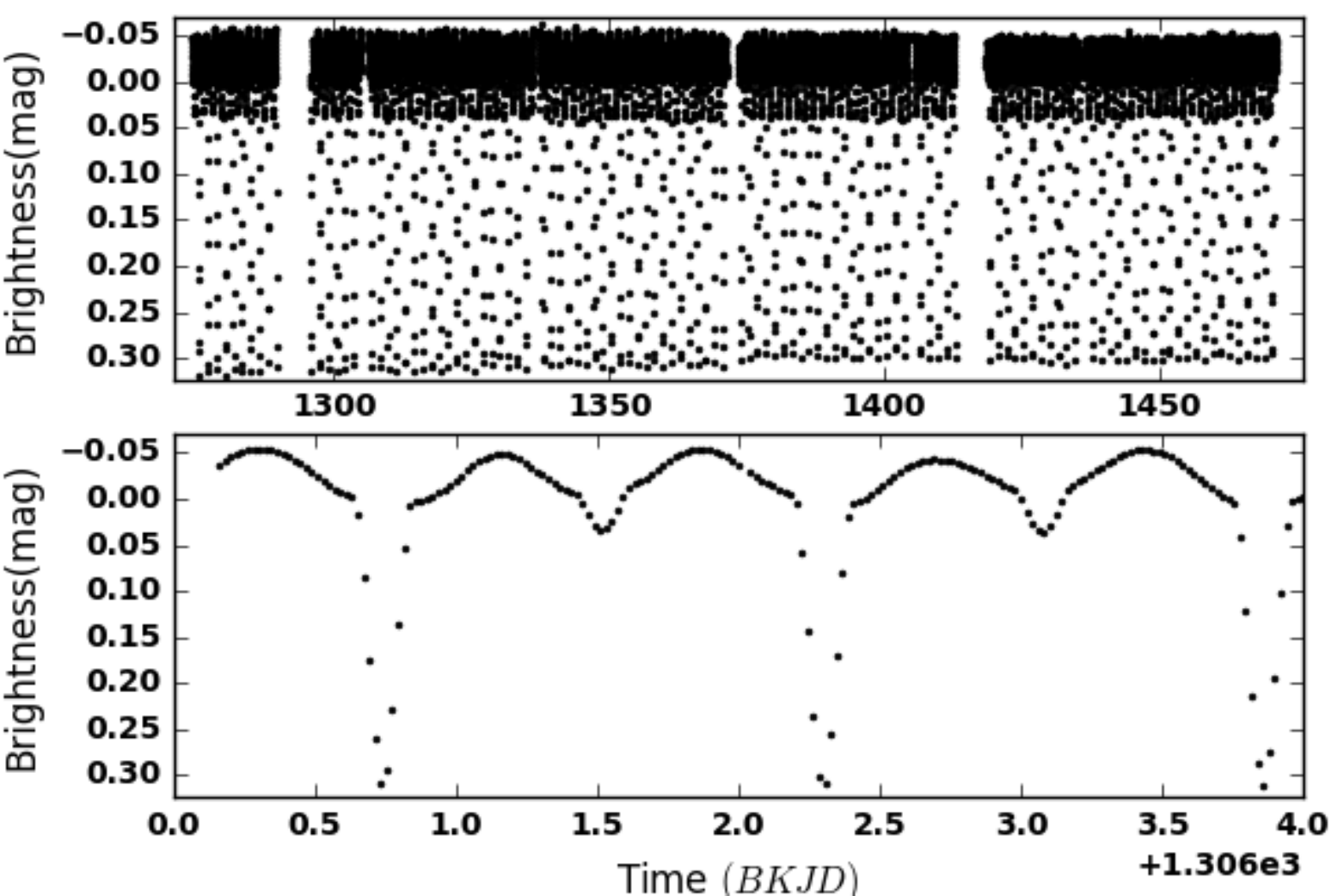}
	\caption{The detrended \textit{Kepler} light curve for the eclipsing binary system KIC 6048106. Top: The full $ Q14 $ and $ Q15 $ light curve, covering 197 days. Bottom: A close view of 10 consecutive days of observations ($BKJD = BJD - 2454833.0$).}
	\label{fig:lc}
\end{figure}
\begin{table}
	\caption{KIC 6048106 observational (photometry) data characteristics.}
	\label{tab:photom_data}
	\begin{tabular}{lcc}
	\hline
	Parameter     & Value in  \href{http://keplerebs.villanova.edu/}{Catalog}\\
	\hline
	Kepler ID     & 6048106\\
	2MASS ID      & 19341402+4123432\\
	RA,~Dec       & +19$^{h}$:34$^{m}$:14.03$^{s}$, +41$\degr$:23$\arcmin $:43.3$\arcsec$\\
	Quarters      & Q14, Q15 \\
	Sampling      & 280   & $\mathrm{\mu Hz}$\\
    Time span     & 196.99661 & d\\
    BJD$_\mathrm{0}$      & 2456107.1603\\
    Rayleigh$~f_\mathrm{res}$& 0.00508 & d$^{-1}$\\
    Period                & 1.559361$\pm$0.000036 & d\\
    Kmag                  & 14.0910\\
	$T_\mathrm{eff}$      & 6777 & K\\
	$\log{g}$             & 4.166 & cgs\\
    $\frac{R}{R_{\sun}}$  & 1.522\\                  
    $\frac{Fe}{H}$        & -0.399 & cgs\\ 
    \hline
    \end{tabular}
\end{table}
KIC 6048106 is a short-period, $P_\mathrm{orb} = 1.56$ d eclipsing binary system. It was discovered by~\citet{Gaulme2010} in their search for pulsations among \textit{Kepler} eclipsing binary stars. KIC 6048106 was listed as a detached binary system with low-frequency pulsations in the range 0.43-3.30~d$^{-1}$. For this target, we used the Long Cadence (LC) light curves from ~\textit{Kepler} with a sampling of 29.42 min. For a typical LC-type \textit{Kepler} observations 270 exposures of 6.02-s are added together to form the ~\citep{Jenkins2010}. Since it takes 372.5 days for \textit{Kepler} to complete one orbit, the terminology "quarter" refers to one quarter of its orbital period i.e 93 days. The first and second quarters are exceptions which cover 10 and 34 days of observations respectively. There are no short cadence data (SC) available for the target.\\
The only \textit{Kepler} light curves available for KIC 6048106 are from quarters 14 and 15. These quarters cover a 197-days observation period from 2012-06-28 to 2013-01-11. The interval comprises 8716 flux-time data points (488 and 414 undefined data points from Q14 and Q15, respectively). There are some significant events which caused some gaps on light curves of both Q14 and Q15. Among those, we can list the failure of \textit{Kepler} second reaction wheel half-way through Q14 month 1 and a safe mode happening in the middle of month 2 of Q15 during which only essential functions were active.\\
Every quarter of the data from \textit{Kepler} has a different flux average than the previous one (because the observations are done using different CCD arrays). They also have different global trends with variable slopes. This means that we need to shift each quarter separately in order to have nearly the same average flux level for different quarters. There also some outliers inside the data, which we need to remove from the light curve. In the case of binary light curves we consider the data points with fluxes larger than 3$\sigma$ of the out-of-eclipse flux level, rather than mean flux level of the light curve, outliers. The observed quarters for our target do not show a large difference in flux level and \textit{Kepler} data reduction pipeline is efficient enough. We can remove the offset and small trends using a polynomial fit to the light curve. Distinguishing the polynomial degree correctly is not as trivial as it seems and a wrong decision can lead us to loose some physical information. We started with removing the eclipses from the light curve and filling the gaps (which were produced from eclipse removal) applying cubic spline fitting. In this case, fitting a polynomial to the reproduced light curve can give us a good idea of the degree of polynomial fit which we need to remove from each quarter. We then inferred the average level flux. Further, we converted the fluxes (including the eclipses) to magnitudes (which can also resolve the amplitude difference between quarters). Finally, we removed the offsets and trends using a polynomial fit to each quarter. We merged the quarters using the average out-of-eclipse brightness level. As the final step in detrending the light curve, we removed the outliers. Figure~\ref{fig:lc} illustrates the detrended light curve for KIC 6048106. The bottom panel is a close view of three orbital cycles (nearly ten days) and shows the brightness level inside and outside the eclipses. We put all the information we have about this target (from KASC and \text{Kepler} eclipsing binary catalog) together in Table~\ref{tab:photom_data}.
\section{Methodology}\label{sec:method}
Our methodology will be based on searching for the binary model that best represents the mean light curve with the aim to report accurately determined stellar and orbital parameters of the  binary system (from the photometry) that are not influenced by the light variations caused by the stellar pulsations. \\
Subsequently, we intended to detect and present stellar pulsation properties in both the low and high-frequency regimes in detail. In order to study both issues in a reliable way, we treated the binarity and pulsation variability behaviours as much as possible separately. As a final step in the observational study of KIC 6048106, we also looked for mutual interactions between these different processes along with the stellar rotation and detected magnetic activity (see later).\\
The main purpose is to acquire a consistent set of accurate constraints for a fine-tuning theoretical approaches of stellar evolution and pulsation models. We describe our approach in three steps:
\subsection{Distinguishing the contribution of the pulsations in the observed light curve} \label{sec:step1}
The idea is that by removing the contribution of global pulsations in the brightness variations of the light curve, we will be able to derive reliable stellar parameters for the components of the binary system. By separately treating the binarity and pulsation, the detected significant independent frequencies will be more reliable to be introduced as pulsation modes. Hence, the first step of our analysis was to define the contribution of the pulsations to light curve.\\
We used consecutive Fourier analyses of the original and prewhitened light curve to find significant frequencies (with stop criterion $ SNR>4 $~\citep{Breger1993}). The noise level and SNRs were calculated within a box size of 1 d$^{-1}$. Our method for Fourier analyses and prewhitening was described in \citet{Degroote2009} in detail (and will explained in paper II. Pulsation study). We applied a multi-parameter sinusoidal fit to all significant frequencies simultaneously, and prewhitened the original light curve with the sum of all respective sinusoidal functions ($m =m_\mathrm{0}+\sum_{n}{A_n\sin(\omega_n t+\phi_n)}$), except for the orbital frequency and its harmonics. In order to identify the orbital frequency and its harmonics we considered the Rayleigh limit as a good approximation of the frequency resolution ($T\geqslant 1/|\nu_{i}-\nu_{j}|,~i\neq j$). The residuals were used as input light curves in the following step (step~\ref{sec:step2}) for binary modelling.
\subsection{Search for the best-fit binary model} \label{sec:step2} 
We used PHOEBE 0.31 (svn 2010-07-08) with an optimization for \textit{Kepler} light curves (finite integration time for two observing cadences) and the selection of the \textit{Kepler} passband.\\
PHOEBE stands for \textit{PHysics Of Eclipsing BinariEs} and is based on the Wilson-Devinney code~\citep{Wilson1971}. The main goal is to simultaneously fit the light and radial velocity curves of binary systems using a wide range of "physical" models, based on the concept of equipotential surfaces for different configurations. A graphical user interface incorporates a wide range of model grids for detached, semi-detached and contact binary systems~\citep{Prsa2005}. The Wilson-Devinney code is based on the inverse problem, which obtains the stellar and orbital parameters from a given light curve. The input light curve should contain the smallest possible systematic and random errors. The code uses differential correction for deriving the sum of the squared deviations $ \chi^\mathrm{2}$ comparing the synthetic light curve and the observed one. The classical Roche model is adopted where it follows the principle of equipotential surfaces. Here both masses are assumed as point sources. Calculation of the gravity darkening coefficients ($gb$) is based on the study of von Zeipel (1924) and Lucy (1967) where the coefficients are $gb=1$ and $\sim0.32$ for stars with radiative and convective envelopes, respectively. For the limb darkening calculations, the code assumes a plane-parallel grey atmosphere approximation to obtain the optical depth. The linear, logarithmic and square-root laws are applied for obtaining the best solution. For the reflection effect, the code uses the geometrical simplification that the irradiated component is a point source. The stellar spots may migrate in longitude following the rotation of the stellar surface, rather than being tied to the coordinate grid~\citep{Wilson2007}.\\
We performed six different runs with PHOEBE with the following sets of models:
\begin{enumerate}
\item A binary system without spots for the full time span of the observations.
\item A binary system with a cold spot on the secondary component for the full time span of the observations.
\item A binary system with a hot spot on the primary component for the full time span of the observations.\\
\\
In Section~\ref{sec:bin_model1} based on visual inspection of the residuals we will show that a single model cannot describe the observed brightness variations during the full time span of the observations. Accordingly, we also considered the following models in Section~\ref{sec:bin_model2}:
\item A binary system with three spots on secondary component for the cycles 817-872 (cycles 828-830 are not observed).
\item A binary system with five spots on secondary component for the cycles 873-916 (cycles 880, 907-909 are not observed).
\item A binary system with two spots on secondary component for the cycles 917-943.
\\
Moreover, in Section~\ref{sec:bin_model1} we will report about an obvious change in the primary eclipse depth. For this reason, we performed two more runs:

\item A binary model with three spots on secondary component for the cycles 817-879 (BKJD From 1274.13986 to 1371.32258).
\item A binary model with two spots on secondary component for the cycles 881-943 (BKJD From 1373.48844 to 1471.136474). 
\end{enumerate}
\subsection{Decoding the Fourier spectrum and study of the stellar pulsations} \label{sec:step3} 
We found that the combination of models (iv), (v), (vi) was the best choice for minimizing the residuals. We concatenated the residuals of the models (iv), (v) and (vi) (section~\ref{sec:step2}). Then we analysed the Fourier spectrum of the residuals after prewhitening for the significant frequencies (with stop criterion $ SNR>4$). The noise level and SNR ratios were calculated within the box size of $1~d^{-1}$. The Fourier spectrum was studied in 3 different regions i.e: $(0.6-3.0)~d^{-1}$, $(3.0-6.0)~d^{-1}$ and $(6.0-24.7)~d^{-1}$. Many significant frequencies with low amplitudes appear in the neighborhood of the dominant frequencies (and also around the harmonics of the orbital frequency in the original Fourier spectrum) among the prewhitened frequencies. According to~\citet{Loumos1978} if the difference between two frequencies is consistent with $ |\nu_{i}-\nu_{j}| \leq 2.5/T$, we cannot resolve them properly. The fact that we prewhitened the observations in two steps (steps~\ref{sec:step1} \&~\ref{sec:step3}), motivated us to consider the mentioned criterion as a safe one to select the most significant frequencies as candidates for pulsations. In the case of two frequencies with difference lower than $ 2.5/T $ we kept the one with the larger amplitude. During our search for g-modes, we looked back to the removed frequencies list (with lower amplitudes) to find a missing frequency based on specific period spacing pattern. \\ 
In this paper, we will discuss the results and present the conclusions concerning the two first steps. This is justified by fact that the search for a consistent binary model throughout the whole time span revealed an additional complexity. Indeed, the long-term magnetic activity of the secondary star challenged us to present different spot activity for different orbital cycles. We studied the ETV curves to find any fingerprint of a long-term trend which also led to a refining of the orbital period. \\
In a subsequent paper (II. Pulsation Study) we intend to present a complete study of the pulsation analysis for this star, after removal of the contributions of the best-fit binary model plus a long-term trend that will be defined in Section~\ref{sec:OC}.
\section{Results}\label{sec:res}
\subsection{Binary modelling-first approach}\label{sec:bin_model1}
\begin{figure}
\includegraphics[width=\columnwidth]{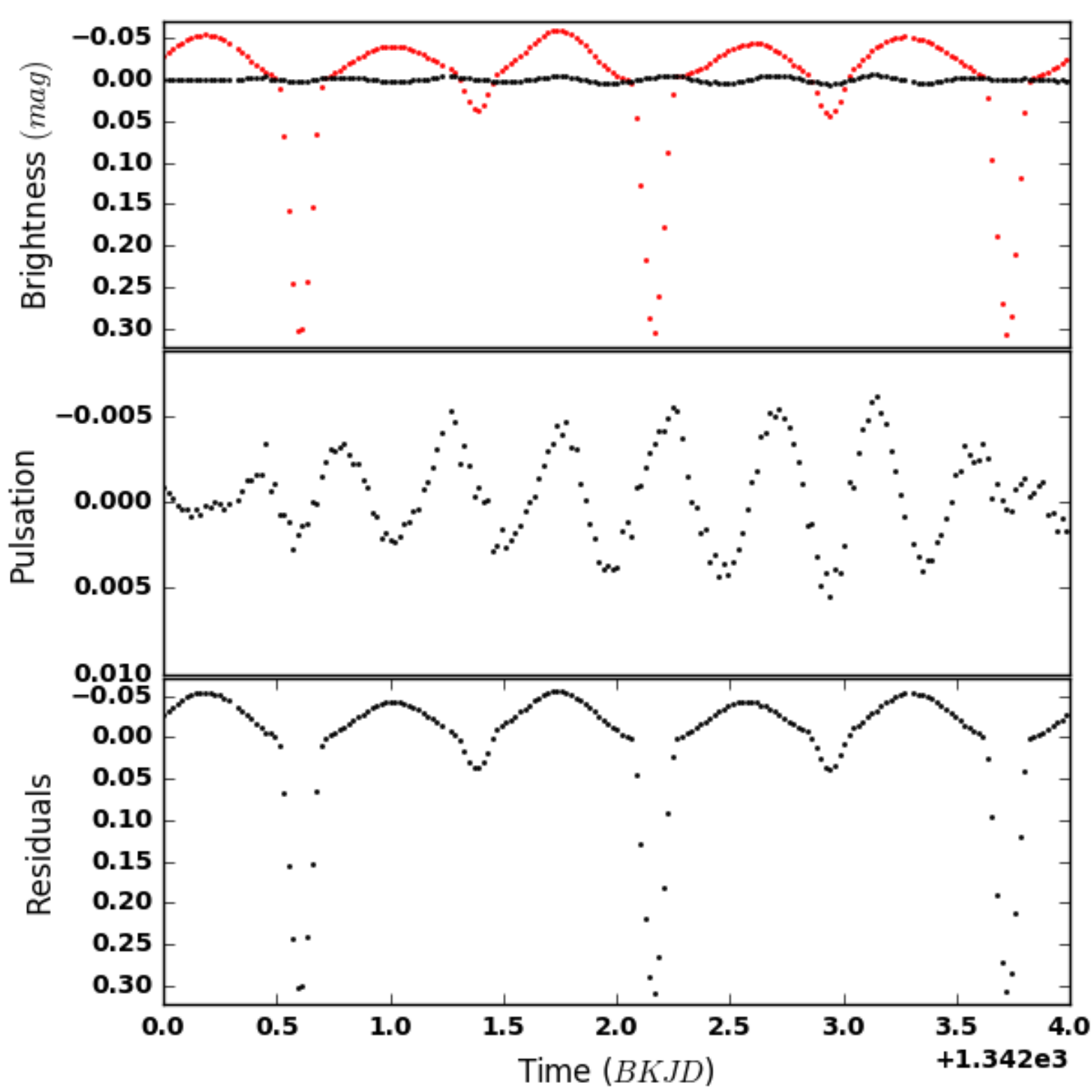}
\caption{Top panel: The original light curve (red curve) compared with the contribution of pulsation in the brightness level (black curve). Middle panel: The effect of pulsations in the light curve, reproduced based on the sum of sinusoidal fits to all prewhitened frequencies excluding orbital frequency and its harmonics. Bottom panel: preliminary binary light curve, which was used to calculate best-fit model.}
\label{fig:modfunc}
\end{figure}
We started our search for a best-fit binary model from the residuals after subtraction of an estimated model representing global stellar pulsations (steps~\ref{sec:step1} and~\ref{sec:step2} in section~\ref{sec:method}). The bottom panel of Figure~\ref{fig:modfunc} presents the section of the light curve prepared for the binary modelling. The amplitude of the light variations due to the global stellar pulsations (middle panel of Figure~\ref{fig:modfunc}) shows its relatively small contribution in the flux variations with respect to the flux variations due to the orbital motion of the system in the original light curve of KIC6048106.\\
For the binary modelling, we used a phase-folded light curve with the reported period of P = 1.559361$\pm$0.000036 d from the \href{http://keplerebs.villanova.edu/}{\textit{Kepler} Eclipsing Binary Catalog}. Considering that the observations from \textit{Kepler} are white-band photometry and the lack of spectroscopic observations, we tried to get the best fit with the least possible free parameters. Some parameters are determined more reliably and precisely from spectroscopy than from photometry. Among those are the eccentricity (e), argument of the periastron ($\omega$) and the mass ratio (q). The importance of fixing these quantities properly is crucial in the absence of spectroscopic observations. The information from the phase and the shape of the eclipses (coupled with $ e\sin{\omega} $ and $ e\cos{\omega} $) along with the least squares minimization method indicates a near-circular orbit. We adjusted the phase and phase difference of the eclipses to be consistent with the observed light curve. Finally, the values of e ($ \simeq $0.01), $\omega$ ($ \simeq $1.64 rad), $\Delta \phi$ (-0.0045) and the phase of the BJD$_\mathrm{0}$ (0.00076) were fixed in order to compute a binary model and a reduced number of free parameters, with the code PHOEBE. \\
\begin{figure}
	\includegraphics[width=\columnwidth]{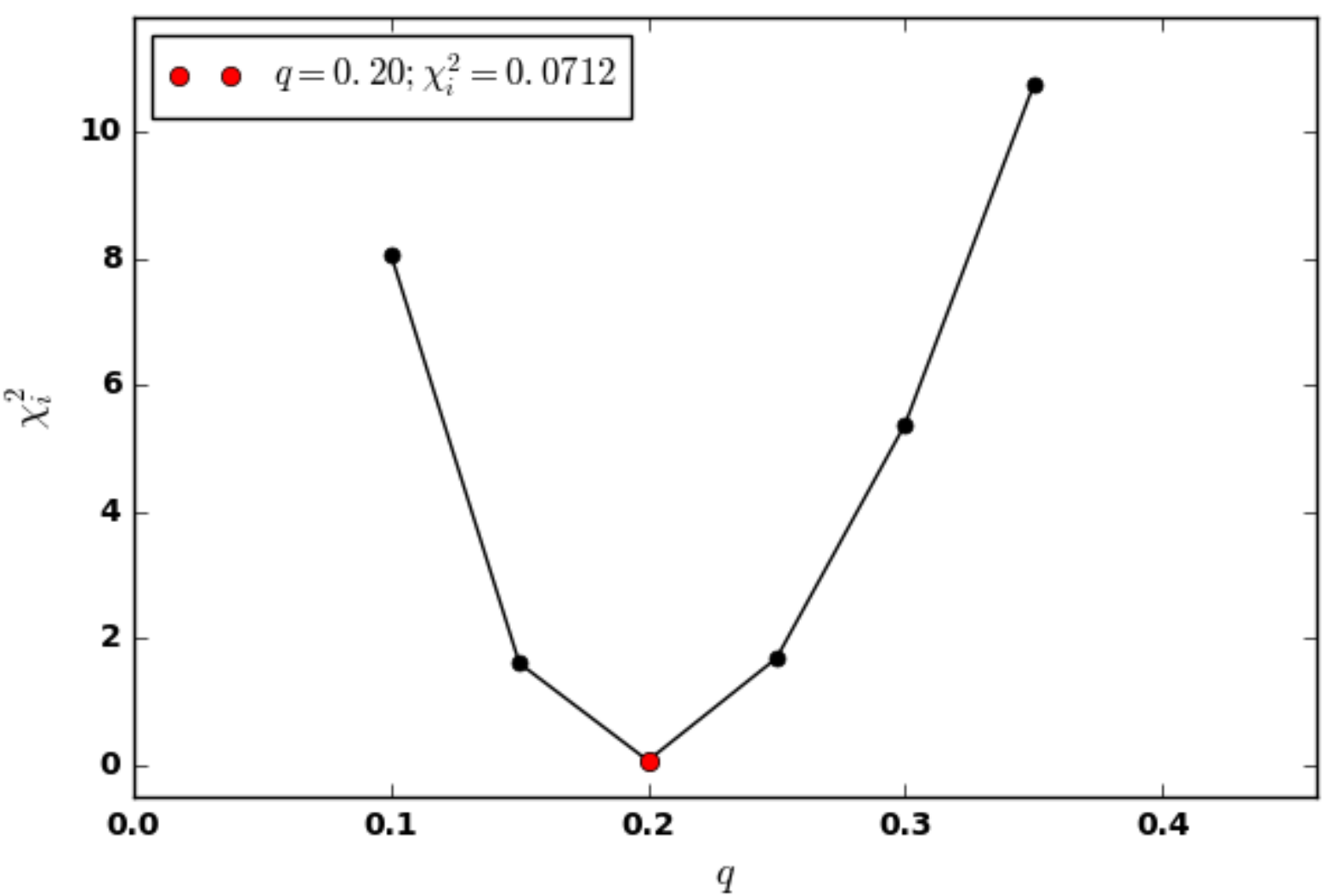}
	\caption{The least squares minimization (LSQ) results to estimate the initial value of mass ratio $q$. Different $\chi_{i}^{2}$ values are plotted for $ q = [0.1-0.35] $ with the step of $ 0.05 $.}
	\label{fig:q}
\end{figure}
In the absence of multicolour photometric observations, we fixed $ T_\mathrm{eff} $ of the primary star around its catalog value of 6777 K for a unique solution with PHOEBE. The surface parameters, albedos and gravity darkenings were fixed to the values for stars with radiative and convective envelopes for primary and secondary components, respectively: $A=1,~0.5$ and $gb = 1,~0.32$. For the limb darkening values, we considered interpolations in the logarithmic limb darkening tables of \citet{Claret2011}. Only one reflection effect was considered. We observed some large asymmetry in the flux level at the wings of the secondary and therefore assumed the presence of spot(s). The binary is an Algol type system where the actual secondary star filled its Roche lobe. Accordingly, we might expect some episodes of mass and angular momentum transfer. However, we ignored this probability and supposed a synchronous rotation for both components ($ F_\mathrm{1,2} = 1 $) during the initial iterations. \\ 
\begin{figure}
\centering
\includegraphics[width=\columnwidth]{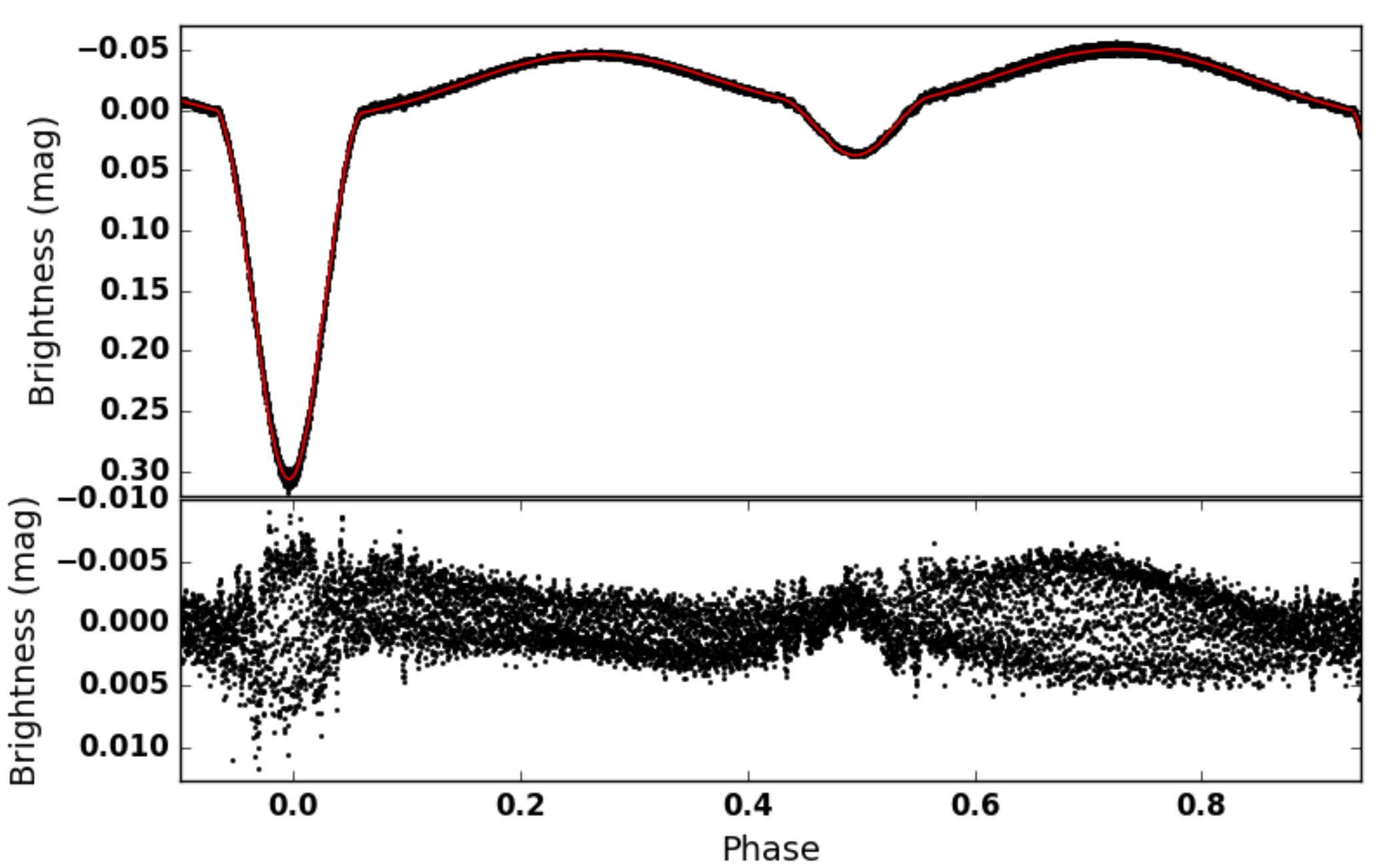}
\caption{Top panel: The best-fit binary model (red) with a cold (Hot) spot on the secondary (primary) component (Tables~\ref{tab:chbinmodel} and \ref{tab:chspot}) and Total time span of the observations (black). Bottom panel: The residuals of the observations after binary modelling. For both models (Hot spot on the primary or cold spot on the secondary star) we have same $ \chi_\mathrm{min}^2 $ and $1\sigma$; i.e. $\chi_\mathrm{min}^2$ = 0.061; 1$\sigma$ = 0.0026 mag. }
\label{fig:chbinmodel}
\end{figure}
\begin{table}
	\caption{The stellar parameters derived with the best-fit model with an either a cold spot on the secondary or hot spot on the primary star. x and y are the linear and non-linear limb darkening coefficients. }
	\label{tab:chbinmodel}
	\begin{tabular}{lcc}
	\hline
 Param.      & Prim.       & Sec. \\
 \hline
 $ M~(\mathrm{M_{\sun}})$ & 1.55$\pm$0.12& 0.33$\pm$0.76  \\   
 $ R~(\mathrm{R_{\sun}})$ & 1.57$\pm$0.12& 1.76$\pm$0.26  \\
 $ T_\mathrm{eff}~(\mathrm{K})  $ & 6791$\pm$114 & 4480$\pm$100   \\
 $ L~(\mathrm{L_{\sun}})$& 4.72$\pm$0.05& 1.03      \\
 $ \log(g)$~(cgs) & 4.23$\pm$0.37& 3.46 $\pm$ 0.59 \\
 $ \Omega       $ & 4.68$\pm$0.03& 2.28     \\
 $ M_\mathrm{bol}      $ & 3.05$\pm$0.01& 4.60     \\
 x$ _\mathrm{bol}      $ & 0.663       & 0.723    \\
 y$ _\mathrm{bol}      $ & 0.194       & 0.074    \\
 x                       & 0.613       & 0.729    \\
 y                       & 0.196       & 0.0963  \\
 $r_\mathrm{pole }     $ & 0.224$\pm$0.015      & 0.235$\pm$0.016 \\
 $r_\mathrm{point}     $ & 0.226$\pm$0.015      & 0.344$\pm$0.033 \\
 $r_\mathrm{side }     $ & 0.225$\pm$0.015      & 0.244$\pm$0.018 \\
 $r_\mathrm{back}      $ & 0.226$\pm$0.015      & 0.278$\pm$0.022 \\
 \hline 
 Binary system   &                  \\
 \hline
 $ M_1+M_2~(\mathrm{M_{\sun}})            $ & 1.87$\pm$0.15       \\
 $ q                  $ & 0.21$\pm$0.05       \\
 $ a~(\mathrm{R}_\mathrm{\sun})$ & 6.98$\pm$0.08                  \\
 $ i\degr          $ & 73.45$\pm$2         \\ 
 \hline
	\end{tabular}
\end{table} 
\begin{table}
	\caption{Physical parameters and spot locations in the best-fit model (Table~\ref{tab:chbinmodel}). The errors obtained by the visual inspection of the effect on the light curve for the co-latitudes, longitudes, radii and temperature scale are in the order of ($ \pm2\degr,~5\degr,~\pm0.5\degr,~0.05 $), respectively.}
	\label{tab:chspot}
	\centering
	\begin{tabular}{lcccc}
	\hline
    Component& Co-latitude  & Longitude &  Radius   &  $\frac{T_\mathrm{spot}}{T_\mathrm{surf}}$ \\
    \hline         
    Primary  & 48$  \degr $           &  110$  \degr $       &  11.5$  \degr $       &   1.03 \\
    or       &                                               \\
    Secondary& 40$  \degr $           &  105$  \degr $       &  11.0$  \degr $       &   0.80 \\
    \hline
	\end{tabular}
\end{table} 
\begin{figure}
\includegraphics[width=\columnwidth]{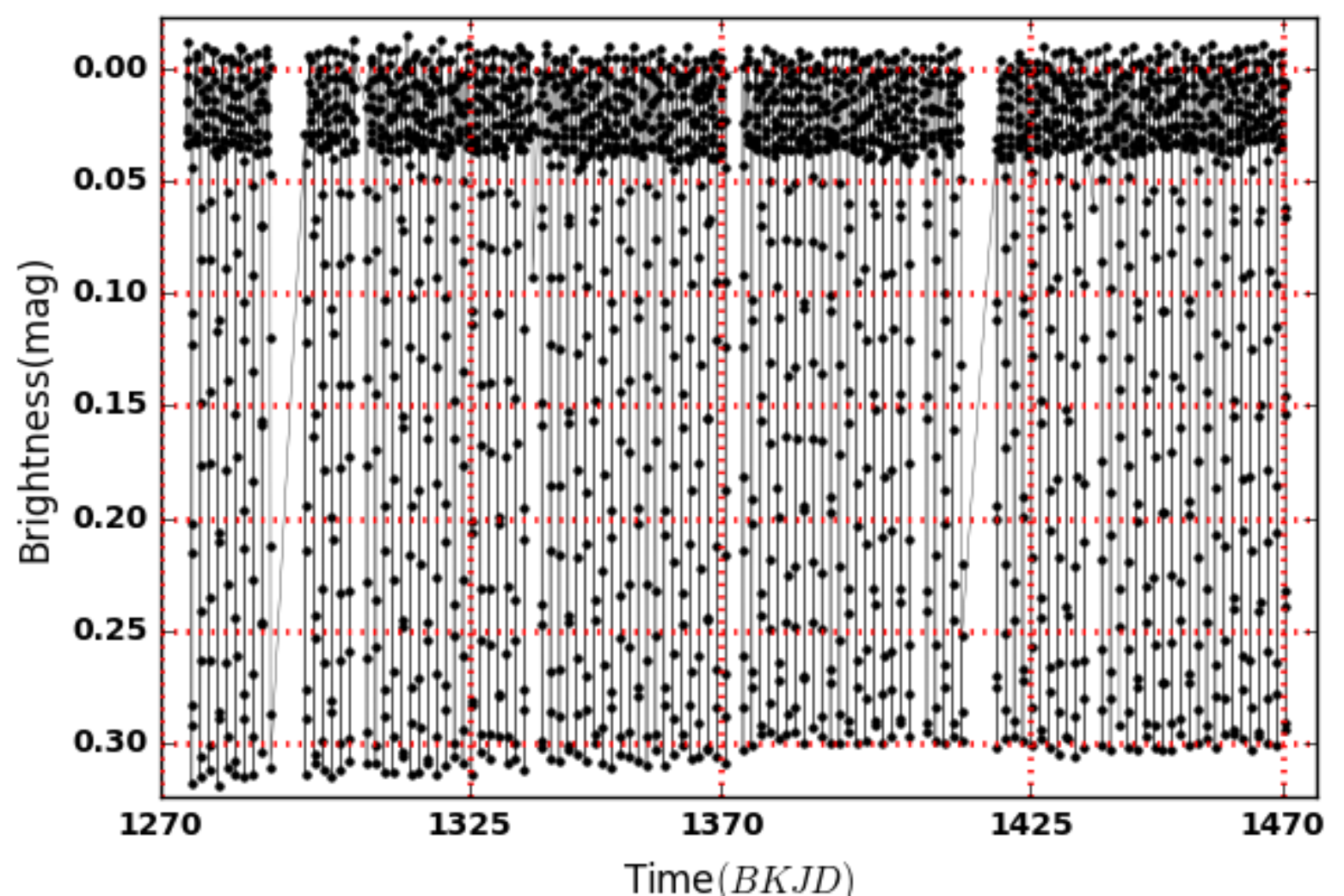}
\caption{The parts of the normalized \textit{Kepler} light curves during primary and secondary eclipse. The depth of the primary eclipse shows an obvious decrease before BKJD $\sim $ 1371.}
\label{fig:eclps}
\end{figure} 
We next estimated the most adequate value for $q$. The $ \chi^2_\mathrm{i} $ values from the least squares minimization are plotted in Figure~\ref{fig:q} for the interval $q$ = [0.05-0.35] and a step of 0.05. The minimum $ \chi^2_\mathrm{min} $ indicates $q$ = 0.2. We then iteratively fitted the following parameters with 4 free parameters, i.e.: $ q,~i,~T_\mathrm{eff2},~\Omega_\mathrm{1} $, where $ \Omega_\mathrm{1} $ refers to the surface potential of the primary star. There is a considerable difference between maximum light at the first and the second quadrature phases (Figure~\ref{fig:chbinmodel}). We can explain such brightness difference either with a hot spot on the primary or a cold spot on the secondary component. The mean standard deviation of the residuals light curve ($ \sigma $) and $\chi^2_\mathrm{min}$ for both models (hot versus cold spot) is nearly the same ($ \chi_\mathrm{min}^2$ = 0.0610;~1$\sigma$ = 0.0026~mag). The stellar parameters of the best-fit binary model are listed in Table~\ref{tab:chbinmodel}. Top panel in Figure~\ref{fig:chbinmodel} shows the best-fit binary model compared with the observed light curve. The residuals are plotted in the bottom panel.\\
Figure~\ref{fig:eclps} shows the parts of the normalized light curve at the times of the primary and the secondary eclipses. Note how the depth of the primary eclipse is decreasing ($ \simeq7\% $) in an interval from start of the observations to BKJD $\sim$ 1371. The decreasing trend stops after BKJD 1371 and we see a nearly constant primary eclipse depth during the remainder of the observations. In order to check if any of the stellar parameters are in charge of this change, we divided the light curve (which is used for binary modeling in this section) in two segments (with a breakpoint at BKJD $\sim$ 1371). The results of the best-fit model to each of the segments show that there is no significant change in any of the stellar parameters except than the (total) luminosity. This indicates that the eclipse depth variation is related to some features on the surface of the star(s).   
\subsection{Binary modelling-second approach}\label{sec:bin_model2}
\begin{figure}
\includegraphics[width=\columnwidth]{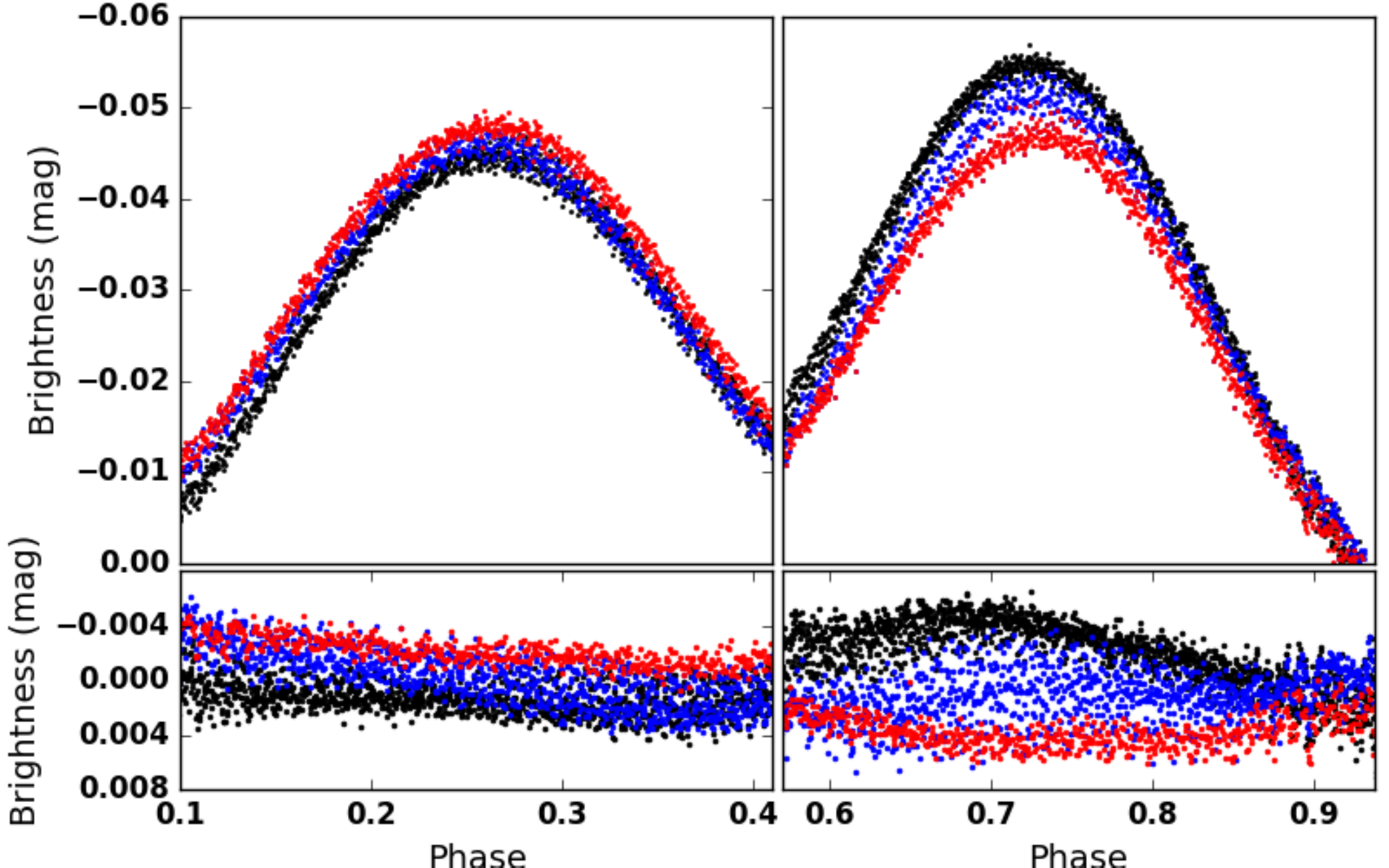}
\caption{A close view of the quadrature phases of KIC 6048106 light curve. The different colours show groups of orbital cycles with low (red), medium (blue) and high brightness modulation at the second quadrature. The figure illustrates that the amplitude of the modulation at the first quadrature is lower than its value at the second quadrature.}
\label{fig:res3cyc}
\end{figure}
\begin{figure}
\includegraphics[width=\columnwidth]{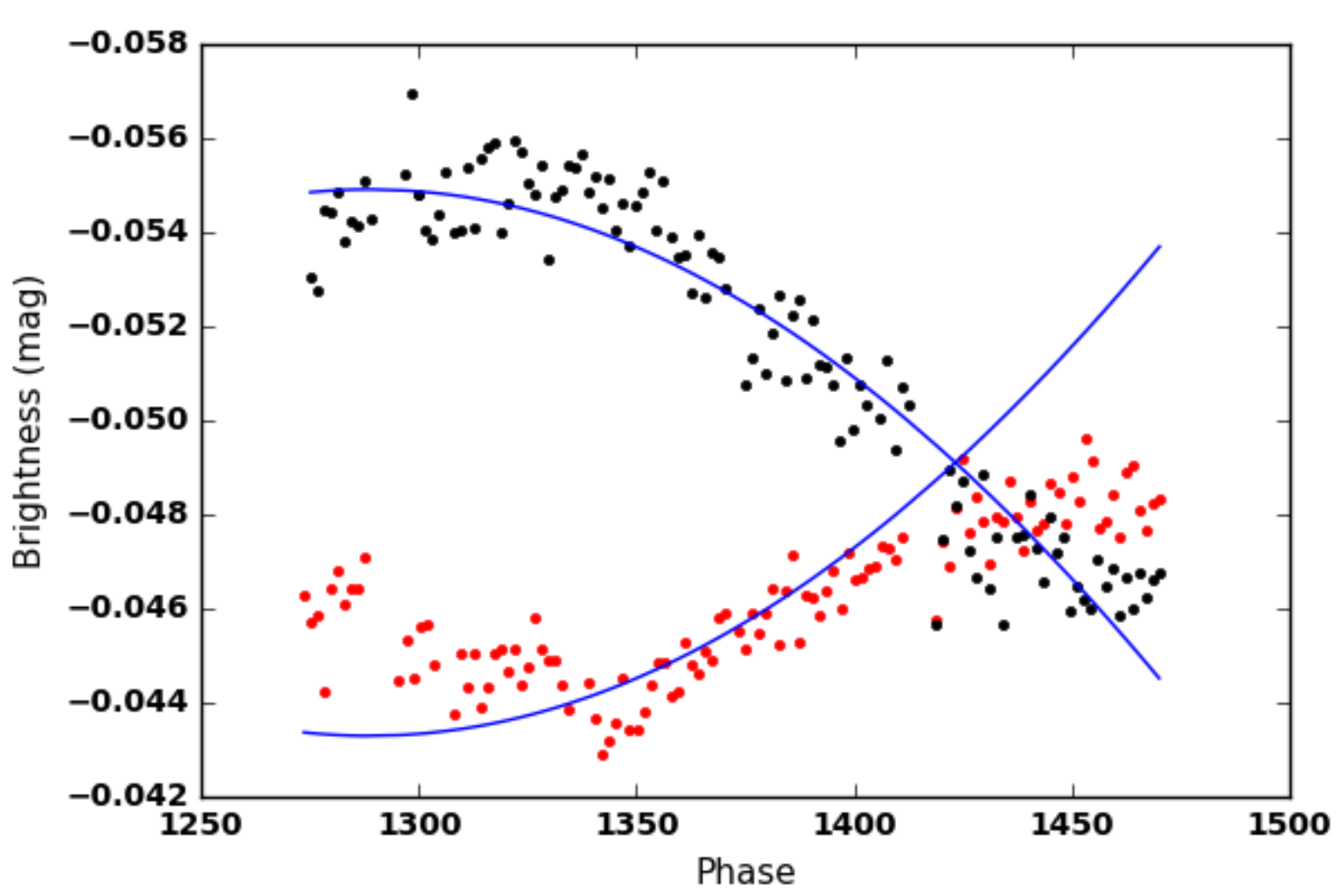}
\caption{The maximum light modulation at first (red) and second (black) quadratures of the light curve. A phase shift between two quadratures has been considered. Blue curve shows a sinusoidal fit to second quadrature which is compared with the modulation at the first quadrature. The frequency of the sinusoidal curve is $ f_\mathrm{mod}$ = 0.003447$\pm$0.000365~d$^{-1} $ ($P_\mathrm{mod}$ = 290$\pm$7 d).}
\label{fig:sinfit}
\end{figure}
\begin{table}
	\caption{The information of the used observations in each state (low, medium, high) and the $\chi_\mathrm{i}^2$ of the best-fit. $N_\mathrm{d}$, $N_\mathrm{cold}$ and $N_\mathrm{hot}$ refer to number of data points in each group of cycles (E), number of cold spots and hot spots, respectively.}
	\label{tab:3dinfo}
	\begin{tabular}{lcccccc}
	\hline
     State & $N_\mathrm{d}$ & E       & $1\sigma_\mathrm{res}$ & $\chi_\mathrm{min}^2$ & $N_\mathrm{cold}$ & $N_\mathrm{hot}$ \\
           &                &         & mag                    &                       &                   &           \\
     \hline       
      High   & 3800   & 817-872 & 0.0015          & 0.0089       &  2         &   1       \\
      Medium & 2912   & 873-916 & 0.0017          & 0.0084       &  3         &   2       \\
      Low    & 1996   & 917-943 & 0.0013          & 0.0033       &  2         &   -       \\
 \hline
 \end{tabular}
\end{table}
\begin{figure*}
\centering
\includegraphics[width=\textwidth]{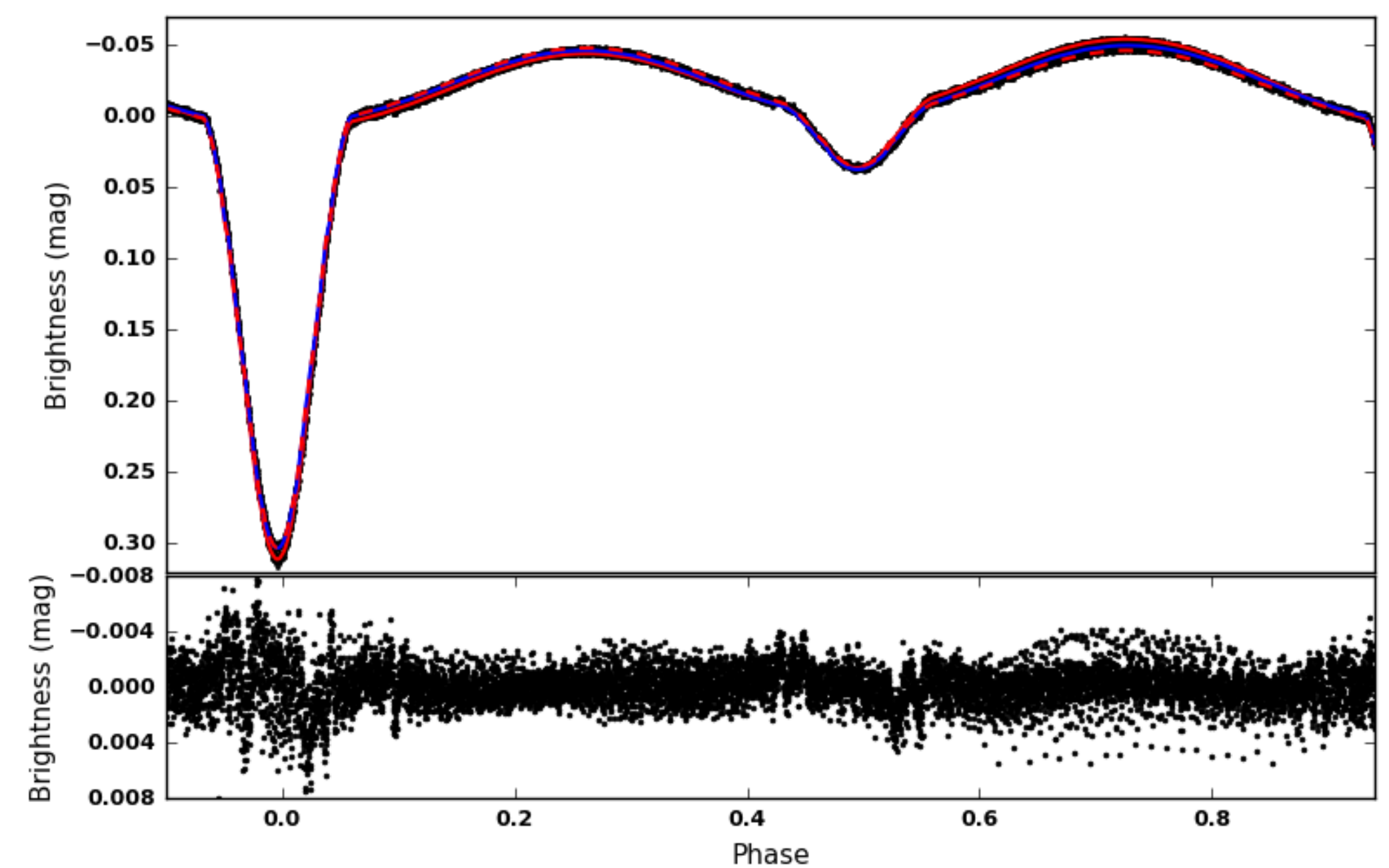}
\caption{The illustration of three best-fit binary models to different group of orbital cycles. 1$\sigma_\mathrm{res} $ refers to the mean standard deviation for the residuals of each binary model. Red solid line: High state 1$\sigma_\mathrm{res}$ = 0.0015 mag. Blue solid line: Medium state $1\sigma_\mathrm{res}$ = 0.0017 mag. Red dashed line: Low state $1\sigma_\mathrm{res}$ = 0.0013 mag. The stellar parameters and the spot information are presented in Tables~\ref{tab:3dstable} and \ref{tab:3mspots}, respectively.}
\label{fig:3dbestfit}
\end{figure*}
\begin{table}
	\caption{The stellar parameters resulting from the best-fit (stable) binary model for three different ranges of orbital cycles. x and y are the linear and non-linear limb darkening coefficients, respectively.}
\label{tab:3dstable}
\begin{tabular}{lcc}
\hline
 Param.        & Prim.     & Sec. \\
 \hline
 $ M~(\mathrm{M_{\sun}})$ & 1.55$\pm$0.11 & 0.33$\pm$0.07  \\   
 $ R~(\mathrm{R_{\sun}})$ & 1.58$\pm$0.12  & 1.78$\pm$0.16  \\
 $ T_\mathrm{eff}~(\mathrm{K})  $ & 7033$\pm$187  & 4522$\pm$103 \\
 $ L~(\mathrm{L_{\sun}})$ & 5.50$\pm$0.05 & 1.14           \\
 $ \log(g)$~(cgs) & 4.24$\pm$0.36 & 3.46 $\pm$0.39 \\
 $ \Omega       $ & 4.69$\pm$0.05 & 2.28           \\
 $ M_\mathrm{bol}      $ & 2.90$\pm$0.01 & 4.55          \\
 x$_\mathrm{bol}       $ & 0.663         & 0.718          \\
 y$_\mathrm{bol}       $ & 0.201         & 0.081          \\
 x                       & 0.613         & 0.722          \\
 y                       & 0.200         & 0.105          \\
 $r_\mathrm{pole }     $ & 0.224$\pm$0.015  & 0.237$\pm$0.016      \\
 $r_\mathrm{point}     $ & 0.226$\pm$0.015  & 0.346$\pm$0.033      \\
 $r_\mathrm{side }     $ & 0.225$\pm$0.015  & 0.246$\pm$0.018      \\
 $r_\mathrm{back}      $ & 0.226$\pm$0.015  & 0.279$\pm$0.022      \\ 
 \hline 
 Binary system   &                  \\
 \hline
 $ M_1+M_2~(\mathrm{M_{\sun}})$ & 1.92$\pm$0.14       \\
 $ q                  $ & 0.22$\pm$0.05               \\
 $ a~(\mathrm{R}_\mathrm{\sun})$ & 7.03$\pm$0.07      \\
 $ i\degr          $ & 73.30$\pm$2                    \\
 \hline
 \end{tabular}
\end{table} 
\begin{table}
	\caption{The physical information and location of spots in three different ranges of orbital cycles (low, medium, high) all on secondary. The errors  obtained by a visual inspection of the effect on the light curve for the co-latitudes, longitudes, radii and temperature scale are of the order of ($ \pm2\degr,~\pm5\degr,~\pm0.5\degr,~\pm0.05 $), respectively.}
	\label{tab:3mspots}
	\begin{tabular}{cccc}
	\hline
 Co-latitude & Longitude &  Radius &  $\frac{T_\mathrm{spot}}{T_\mathrm{surf}}$\\
 \hline  
         & High state  &\\ 
 \hline                                                  
 45$  \degr $           &  70$  \degr $     &  13.5$ \degr $     &  0.78      \\
 35$  \degr $           &  175$ \degr $     &  14.0$ \degr $     &  0.75      \\
 47$  \degr $           &  290$ \degr $     &  11.5$ \degr $     &  1.05      \\
 \hline  
       & Medium state    &\\ 
 \hline                                                  
 45$  \degr $           &  125$ \degr $     &  9.0$ \degr $      &  0.95      \\
 35$  \degr $           &  225$ \degr $     &  8.0$ \degr $      &  0.85       \\
 47$  \degr $           &  330$ \degr $     &  8.0$ \degr $      &  1.05      \\
 28$  \degr $           &   45$ \degr $     & 10.0$ \degr $      &  0.65       \\
 30$  \degr $           &  290$ \degr $     &  5.0$ \degr $      &  1.10       \\
 \hline 
         &  Low state    &\\ 
 \hline    
  45$  \degr $         &  210$  \degr $     &   8.0$\degr $      &  0.85       \\
  35$  \degr $         &  300$  \degr $     &   8.5$\degr $      &  0.95       \\
\hline
\end{tabular}
\end{table}
The residuals of the best-fit model (Section ~\ref{sec:bin_model1}) are significantly larger at the quadrature phases than other phases (in particular the second). Considering the variation of the residual brightness at the second quadrature, we divided the full time span of the residuals from Section~\ref{sec:step1} into three groups of orbital cycles. Figure~\ref{fig:res3cyc} shows the cyclic variation for the three groups: they correspond to a state of low (red dashed line), medium (blue solid line) and high (red solid line) brightness modulation of the second quadrature phase. The residuals of the best-fit binary model are illustrated in the bottom panel. The number of cycles observed in the low state is smaller than the number of cycles in the high and medium states. Figure~\ref{fig:sinfit} displays the maximum magnitude at both quadrature phases for each orbital cycle. The data points for the first quadrature (red) were shifted in time to be comparable to the data points at the second quadrature (black). The plot shows a long-term quasi-sinusoidal light variation only half of which was observed by \textit{Kepler}. A sinusoidal fit to the maximum magnitudes at phase 0.75 shows that the period of this modulation is 290 d ($ f_\mathrm{mod} = 0.003447\pm0.000365~d^{-1} $). Furthermore, a Fourier analysis of the same data resulted in about the same frequency with an amplitude of nearly 0.0045$\pm$0.0004 mag. The amplitude of the brightness modulation at the first quadrature is lower than at the second quadrature and that is why the fit at phase 0.75 can not define well the observed modulation at phase 0.25. \\
In this second approach and to minimize the residuals at the maximum brightness phases, we looked for the best-fit binary model for each of three states (low, medium and high). The same method and similar iterations as applied in Section~\ref{sec:bin_model1} was next used to find a unique binary model (without any spots) for fitting all three states simultaneously. A unique synthetic model was found to be consistent with all three states (which we will call the stable model). By adding different spots to this stable model, we obtained a much improved fit (lower and smooth residuals) describing the observed light curve very well. Figure~\ref{fig:3dbestfit} illustrates the stable model with the different spot models included. Top panel in Figure~\ref{fig:3dbestfit} shows the observed light curve (black) and the best-fit binary models to each of the low (red dashed line), medium (blue solid line) and high (red solid line) state. The average mean standard deviation of the residuals ($\sigma \simeq $ 0.0015 mag) is lowered by as much as 40\% compared to the value in Section~\ref{sec:bin_model1} ($\sigma \simeq $0.061 mag, see Figure~\ref{fig:chbinmodel}). More information about the observations defining each light curve segment is presented in Table~\ref{tab:3dinfo}. It includes the number of data points in each light curve segment ($N_\mathrm{d}$), $\sigma$ of the residuals, $ \chi_\mathrm{min}^2$ of the best-fit and the number of cold ($N_\mathrm{cold}$) and hot spots ($N_\mathrm{hot}$) on the secondary star for each of the three states along with the orbital cycle numbers of each group (E). The smallest value $ \chi_\mathrm{i}^2$ = 0.0033 corresponds to the light curve in the lowest brightness state. The stellar parameters describing the stable binary configuration are listed in Table~\ref{tab:3dstable}. The spot information for each state is presented in Table~\ref{tab:3mspots}. An interesting result, which is presented in Table~\ref{tab:3state_comp}, is that the change in luminosity, in three states, must probably be due to some surface phenomenon, since it is not accompanied by a temperature nor by a radius change of the stars. The system configuration and the spot locations at phases (0.0, 0.25, 0.5, 0.75) are illustrated in Figure~\ref{fig:meshplot}. The top panel shows both components in the lower activity state. The middle and bottom panels refer to the medium and high activity cycles, respectively. For the error estimations, in the absence of any spectroscopic information we used $ \chi_{i}^2$ and the laws of the propagation of the errors.\\
It is good to remind here that the first cycles of the light curve are related to the highest state of the activity and the last cycles of the light curve are mostly related to the lowest activity. One can easily verify that the depth of the primary eclipse is nearly constant during the low state. We know that magnetic field can affect the angular momentum distribution in the outer stellar layers. "These modulations can be explained by the gravitational coupling of the orbit to variations in the shape of a magnetically active star in the system~\citep{Applegate1992}". A change in the shape of the active star (the secondary component of the system) can affect the area (from the primary star) which is eclipsed. Consequently, it can change the level of the minimum brightness during the eclipse (here the primary eclipse). 
\subsection{Analysis of ETV curve}\label{sec:OC}
\begin{figure}
\includegraphics[width=\columnwidth]{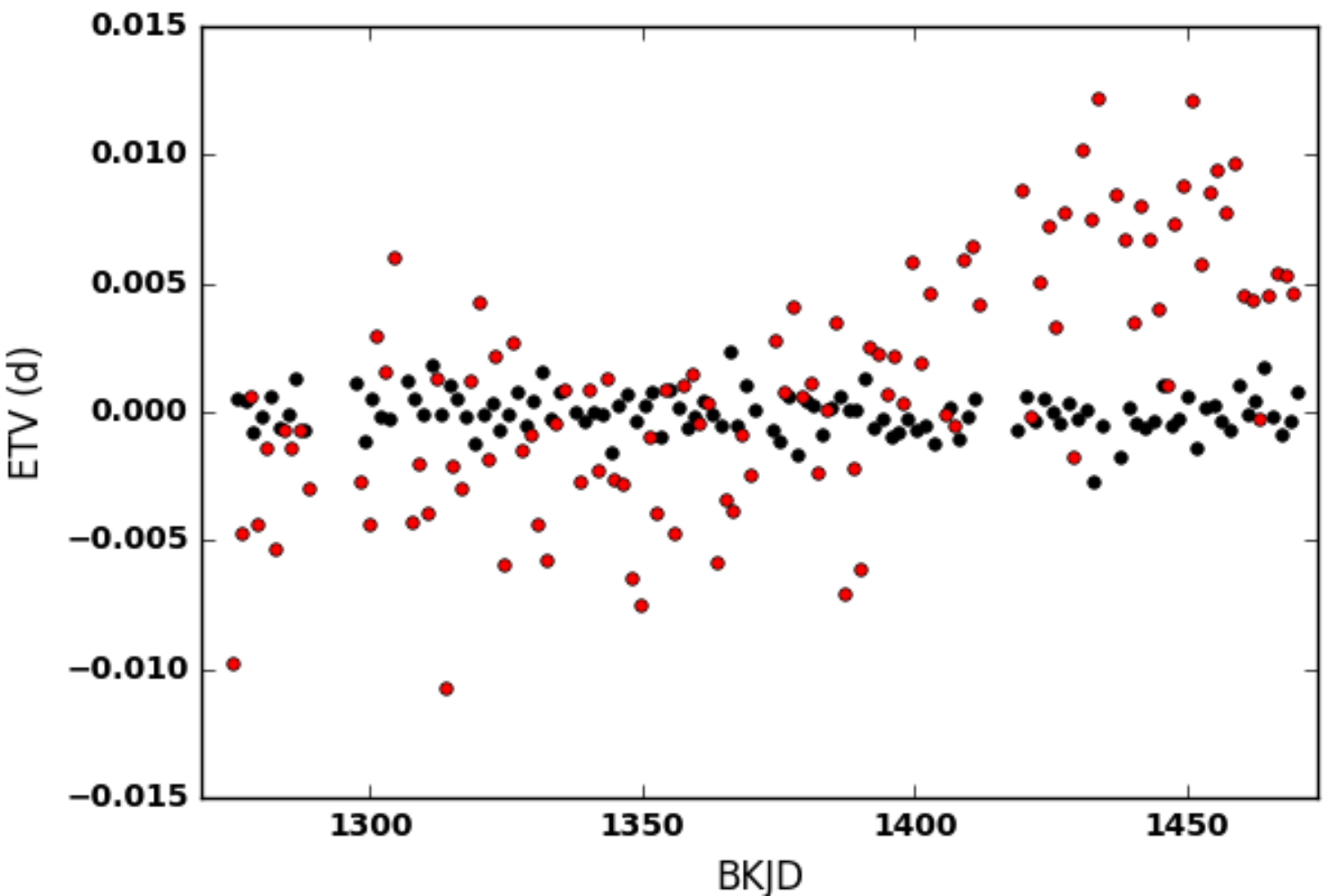}
\caption{The ETV curve of the first (black) and second (red) minima from \textit{Kepler} eclipsing binary catalog. The ETVs of the secondary minima show a clear upward trend.}
\label{fig:oc_kplr}
\end{figure}
\begin{figure}
\includegraphics[width=\columnwidth]{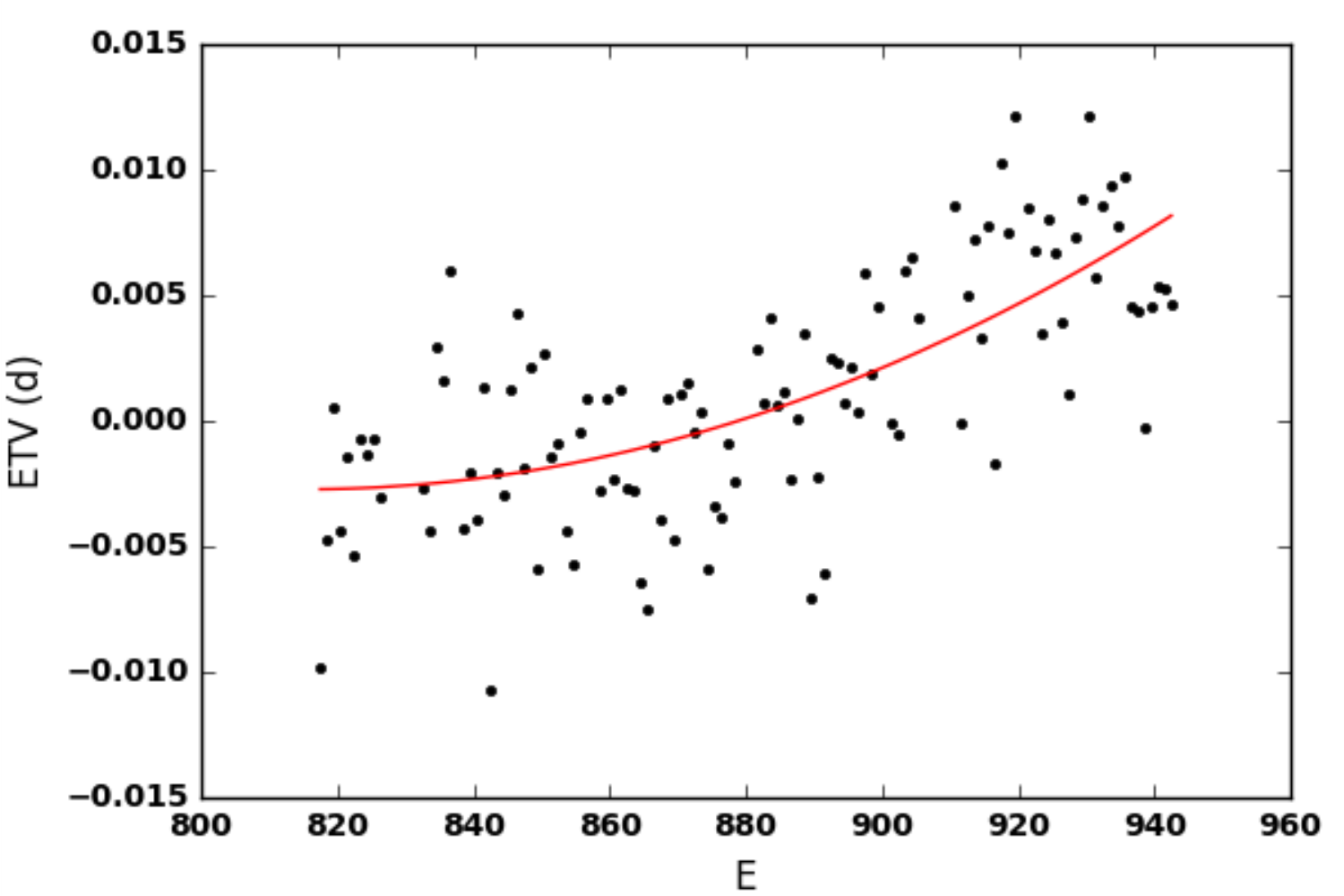}
\caption{A quadratic fit to the ETV curve of the secondary star. E refers to the orbital cycle number. The amplitude of the modulation is $ \Delta P~=~1.34\times 10^{-6} $ d.}
\label{fig:oc_quadfit}
\end{figure}
\begin{figure}
\includegraphics[width=\columnwidth]{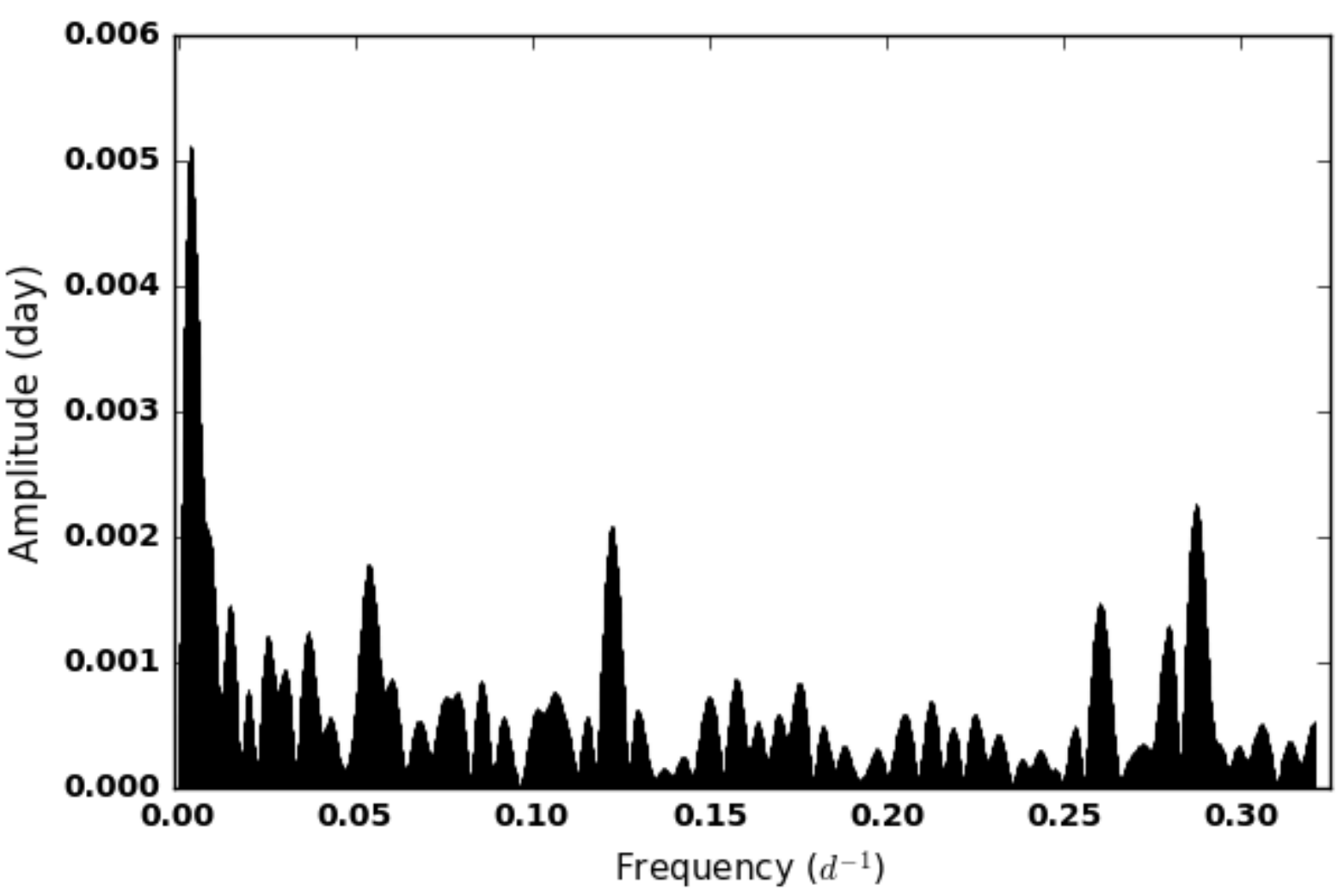}
\caption{The Fourier spectrum of the secondary ETV. The most dominant frequency is $ f_\mathrm{1}$ = 0.00359$\pm0.00009$ d$^{-1}$, which shows a long-term modulation about $ P_\mathrm{mod}~=~278\pm7$ d (Table~\ref{tab:oc_fourier}).}
\label{fig:oc_fourier}
\end{figure}
\begin{table*}
	\caption{The frequency analysis of the ETV curve for the secondary minimum.}
	\label{tab:oc_fourier}
	\begin{tabular}{lccccccc}
	\hline
	$f_{n}$ & $freq$  & $A$     & $\phi$   & $\epsilon_\mathrm{freq}$ & $P_\mathrm{mod}$   & $\dfrac{f_\mathrm{n}}{f_\mathrm{1}}$ & SNR\\
	        & d$^{-1}$& d       &          & d$^{-1}$                 &   d                &                                      & \small{box size (2 d$^{-1}$)}\\
    \hline
    $f_\mathrm{1}$ & 0.00359 & 0.004599 & 0.009789 &  0.00008          &  278.46&   1.0  &   26.17 \\
    $f_\mathrm{2}$ & 0.05310 & 0.002463 & 0.074913 &  0.00016          &  18.83 &   1.48 &   14.08 \\
    $f_\mathrm{3}$ & 1.22358 & 0.002296 & 0.814878 &  0.00017          &  8.17  &   3.41 &   12.95 \\
    $f_\mathrm{4}$ & 2.87041 & 0.002079 & 0.601506 &  0.00019          &  3.48  &   7.99 &   11.71 \\
    $f_\mathrm{5}$ & 2.59850 & 0.001332 & 0.340696 &  0.00028          &  3.85  &   7.24 &   7.48  \\
    $f_\mathrm{6}$ & 2.76523 & 0.001071 & 0.478641 &  0.00037          &  3.62  &   7.70 &   5.99  \\
    $f_\mathrm{7}$ & 1.57757 & 0.000698 & 0.645574 &  0.00057          &  6.34  &   4.39 &   3.82  \\
  \hline
  \end{tabular}
\end{table*} 
Figure~\ref{fig:oc_kplr} shows the difference between the times of light minimum and the ephemeris found in Kepler Eclipsing Binary Catalog: 
\begin{equation}
\label{eq:ephemeri1}
 T_\mathrm{min} = 2454833.0 + 1.556931(\pm0.000036)E 
\end{equation}
E in equation~\ref{eq:ephemeri1} refers to the number of orbital cycle. In the case of Algol binary systems, the magnetic activity of one of the components can cause period modulations typically of the order of $ \frac{\Delta P}{P}\sim10^{-5} $ on a time scale of decades or even longer~\citep{Applegate1992}. These modulations can show off in the ETVs of the active star.\\
The ETV values for the primary star show a constant trend around zero. On the other hand, ETV values for the secondary star are indicating a positive slope. In order to estimate the amplitude of the modulation, we followed the \citet{Kalimeris1994} method which considers the O-C variations as a polynomial function. Figure~\ref{fig:oc_quadfit} illustrates the O-C curve for the secondary star, fitted with a quadratic function: $ETV~=~c_\mathrm{0}+~c_\mathrm{1}~E+~c_\mathrm{2}~E^2$.
The coefficients are:\\
$c_\mathrm{0}~=~0.44\pm0.21~d,~c_\mathrm{1}~=~-0.0011\pm0.0005~d$,\\
$c_\mathrm{2}~=~6.72\times 10^{-7}\pm2.75\times 10^{-7}~\mathrm{d}$, corresponding to a period modulation of the order of $\frac{\Delta P}{P}~=~8.63\times 10^{-7}\pm1.76\times10^{-7}$.\\
We detected five significant frequencies from a Fourier analysis of the O-C curve (figure~\ref{fig:oc_fourier}). The SNR calculations were done inside a 2~d$ ^{-1} $ box. The most dominant frequency is found to be $f_\mathrm{1}$ = 0.00359$\pm$0.00009~d$^{-1}$ with the periodicity of $ P_\mathrm{mod} = 278\pm7$ d. This value is in excellent agreement with the periodicity of the long-term light modulation from Section~\ref{sec:bin_model2} ($ P_\mathrm{mod} = 290\pm7 d$) caused by the presence of spots on the stellar surface. Table~\ref{tab:oc_fourier} presents the information of the additional significant frequencies, also detected from the Fourier analysis of the ETV curve, their amplitude is however much lower. Among these frequencies, $f_\mathrm{4}$ = 2.87 d$^{-1}$ is the 8th harmonic of modulation frequency.
\section{Summary and discussion}
\label{sec:con}
We presented a detailed study of KIC 6048106, a \textit{Kepler} short-period eclipsing binary with the mean Kp magnitude of 14.1 mag. The photometric data comes from the Q14 and Q15 quarters with LC sampling. For the orbital period, we considered the period of the Catalogue of the \textit{Kepler} Eclipsing Binaries, $ P_\mathrm{orb}$ = 1.559361$\pm$0.000036 d. Our aims are twofold: (1) to accurately determine the fundamental parameters of both components of this semi-detached eclipsing system and (2) to study and characterize the pulsation modes of the residual light curve (after removal of the appropriate binary model). We treated the light variations due to binarity and the pulsations separately. In order to derive the fundamental stellar parameters from the light curve with PHOEBE, we first removed the approximated contribution of the pulsations in the light curve of the system by subtracting a multi-parameter fit based on all significant frequencies (SNR>4 d$^{-1}$) from a fast Fourier analysis but excluding the known orbital frequency and its harmonics. Due to the significant light amplitude at two quadrature phases, we added a small cold spot ($\frac{T_\mathrm{spot}}{T_\mathrm{surf}}$ = 0.80$\pm$0.05) on the surface of the secondary star. The same effect on the light curve could also be explained by the presence of a hot spot ($\frac{T_\mathrm{spot}}{T_\mathrm{surf}}$ = 1.03$\pm$0.05) on the primary star. In both cases, the stellar parameters remained constant within the 1$\sigma$-errorbars. The primary star is an intermediate-type ($\sim $F5) main-sequence star. Furthermore, we detected a long-term cyclic variation, causing brightness fluctuations in the residuals of the best-fit binary model. The variations of the maximum light could be modeled by a variation with a period $ P_\mathrm{mod}$ = 290$\pm$7 d. We next were able to derive a consistent binary model which describes very well the observations grouped in three time intervals according to a low, medium and high level of maximum light at the second quadrature. We achieved this by including different spot models which may represent various states of activity, namely a low, a medium and a high state. We considered all the spots on the surface of the secondary star (the evolved component) as it is a cool evolved star (in our case a subgiant $\sim$K5) with a convective envelope and it resides in an Algol-type binary system~\citep{Berdyugina2005}. The fit to the observations with the lowest surface activity has the lowest $\chi_\mathrm{i}^2$ compared with the other two states. The mean standard deviation of the residuals has been reduced by 40\% by including a varying number of spots in the binary modelling. The stellar parameters thus reveal a semi-detached (Algol-type) eclipsing binary system with a near-circular orbit and assumed synchronized fast rotation (with estimated $v\sin{i}$ values for the primary and secondary components respectively around 44 and 50 $\mathrm{kms}^{-1}$ from derived stellar parameters), and magnetic activity of the secondary component due to active (and possibly migrating) spots. The inclination of the orbital plane is found to be $i = 73.30\degr\pm2\degr$. The mass ratio of the components equals $q$ = 0.21$\pm$ 0.05. The radii of the primary and secondary components are $ R_\mathrm{1},~R_\mathrm{2}$ = 1.56 $\pm$0.1, 1.78$\pm$0.30 $R_{\sun} $ and their effective temperatures are $ T_\mathrm{eff}$ = 7050$\pm$106,~4519$\pm$118 K, respectively. \\
The study of the ETV curve from the \textit{Kepler} ETVs shows a constant trend for the primary and a variation with a positive slope for the secondary component. A quadratic fit to the latter ETV values shows a period variation up to $\frac{dP}{dE} = 1.34\times10^{-6}$ d and $\frac{\Delta P}{P}= 8.63\times10^{-7}\pm1.76\times10^{-7} $. Accordingly $P_\mathrm{orb}$ (from \textit{Kepler} Eclipsing Binary Catalog) should be corrected to $P_\mathrm{orb}$ = 1.5582645$\pm$0.0005203 d. Thus the new ephemeris is:
\begin{equation}
T_\mathrm{min}~=~2454833.0~+~1.5582645~E~+~6.72\times 10^{-7}~E^2.
\end{equation}
Moreover, we inferred a magnetic activity cycle with a period of $ P_\mathrm{mod} = 278\pm7$ d (explained by the \citet {Applegate1992} mechanism). This value is in good agreement with the long-term period of spot activity detected through the process of the binary light curve modelling process. Among other frequencies detected from prewhitening of the O-C variations with a sum of sinusoidal functions, $ f_\mathrm{4} = 2.87041\pm0.00019~d^{-1}$. It might correspond to the 8th harmonic (SNR $\simeq 12$) of the modulation frequency ($f_\mathrm{1}$ = 0.00359$\pm$0.00008 d$^{-1}$). This can be explained by the migration of spots which will cause different brightness levels at the quadratures of different cycles. We also reported a continuous modulation in the depth of the primary eclipse before BKJD 1371. A set of the stellar parameters (specifically effective temperature and radius of the stars) derived from the best-fit binary model to the part before BKJD 1371, compared with the stellar parameters to the part of the light curve after BKJD 1371 shows no significant variation. It's just the total luminosity that is varies before and after the break-point (BKJD 1371). This confirms that the variation is happened because of some surface phenomena like spots.\\
In paper II, we aim to describe the results of a detailed study of the hybrid pulsation associated to the primary component of this very interesting binary system.
\section{Acknowledgements}
All of the data presented in this paper were obtained from the Mikulski Archive for Space Telescopes (MAST). STScI is operated by the Association of Universities for Research in Astronomy, Inc., under NASA contract NAS5-26555. Support for MAST for non-HST data is provided by the NASA Office of Space Science via grant NNX09AF08G and by other grants and contracts.\\
Special thanks to \textit{Rahim Heidarnia} from RIAAM for the very useful discussions on binary modelling and \textit{Paul Van Cauteren} from the Humain observatory (ROB) for his continuous support. We gratefully acknowledge a 1-month visitor's grant of the non-profit association "ASBL-VZW Dynamics of the Solar System", and the hospitality of the Royal Observatory of Belgium for the first author.

\appendix
\section{The Binary Configuration of KIC 6048106}
\begin{figure}
\centering
\includegraphics[width=\textwidth]{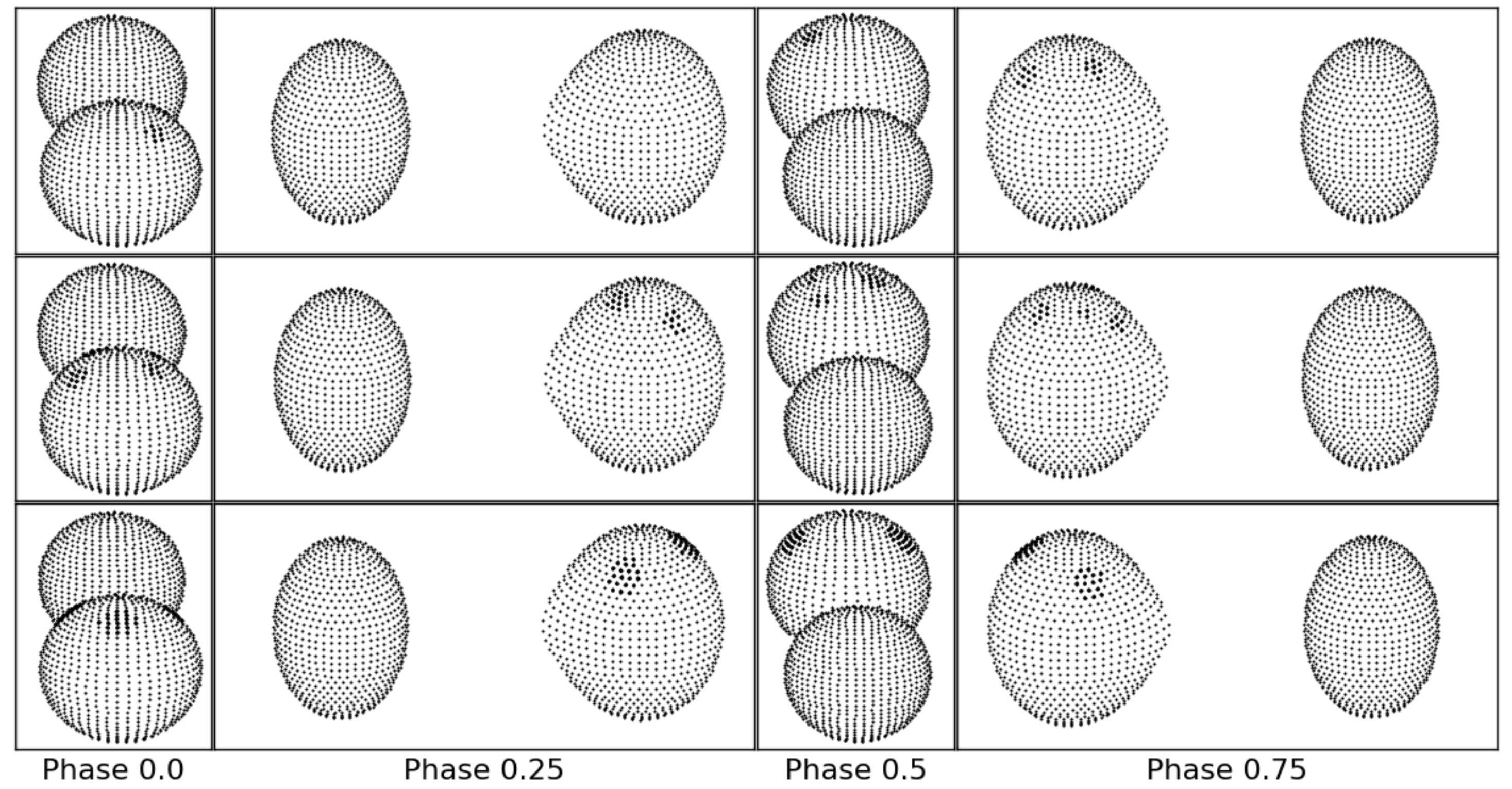}
\caption{The binary configuration and the spot locations for each activity state in four different phases (0.0, 0.25, 0.5, 0.75). Top panel: Low state. Middle panel: Medium state. Bottom panel: High state.}
\label{fig:meshplot}
\end{figure}

Table~\ref{tab:3state_comp} shows the fundamental stellar parameters that were derived from the best-fit binary model to each of the three segments of the light curve with '\textit{High}', '\textit{Medium}', and '\textit{Low}' activity. Figure~\ref{fig:meshplot} shows 
shows the binary system configuration during the eclipse and the quadrature phases, for different sates of activity (Table~\ref{tab:3mspots}). We used the meshplot tool of PHOEBE to prepare this plot. The bold points on the surface of both stars are showing the spots (sizes and locations).
\begin{table}
	\caption{The stellar parameters resulting from the best-fit binary model to three different segments of the KIC 6048106 observed light curve.}
\label{tab:3state_comp}
\begin{tabular}{lccc}
\hline                                                                                 
 Param.                               & High             &   Medium        &    Low    \\
 \hline
 $ L_\mathrm{1}~(\mathrm{L_{\sun}})$  & 5.40 $\pm$0.05   & 5.33 $\pm$0.05  &  5.57 $\pm$0.05          \\
 $ L_\mathrm{2}~(\mathrm{L_{\sun}})$  & 1.18             & 1.10            &  1.17                    \\                                                                               
 $ M_\mathrm{bol1}$                   & 2.91 $\pm$0.01   & 2.92 $\pm$0.01  &  2.87 $\pm$0.01          \\
 $ M_\mathrm{bol2}$                   & 4.51             & 4.56            &  3.46                \\
 $ T_\mathrm{eff1}~(\mathrm{K})  $    & 7044 $\pm$107    & 7001 $\pm$107   &  7054 $\pm$106          \\
 $ T_\mathrm{eff2}~(\mathrm{K})  $    & 4546 $\pm$103    & 4510 $\pm$103   &  4510 $\pm$103          \\
 $ M_\mathrm{1}~(\mathrm{M_{\sun}})$  & 1.54 $\pm$0.12   & 1.55 $\pm$0.12  &  1.56 $\pm$0.12          \\ 
 $ M_\mathrm{2}~(\mathrm{M_{\sun}})$  & 0.34 $\pm$0.07   & 0.33 $\pm$0.07  &  0.33 $\pm$0.04          \\  
 $ R_\mathrm{1}~(\mathrm{R_{\sun}})$  & 1.56 $\pm$0.12   & 1.57 $\pm$0.12  &  1.58 $\pm$0.12          \\
 $ R_\mathrm{2}~(\mathrm{R_{\sun}})$  & 1.79 $\pm$0.16   & 1.78 $\pm$0.16  &  1.77 $\pm$0.16          \\
 $ \Omega_\mathrm{1}$                 & 4.73 $\pm$0.05   & 4.68 $\pm$0.05  &  4.65 $\pm$0.05          \\
 $ \Omega_\mathrm{2}$                 & 2.30             & 2.28            &  2.27               \\
 $ \log(g_\mathrm{1})$~(cgs)          & 4.24 $\pm$0.36   & 4.24 $\pm$0.36  &  4.23 $\pm$0.36          \\
 $ \log(g_\mathrm{2})$~(cgs)          & 3.47 $\pm$0.39   & 3.46 $\pm$0.39  &  3.46 $\pm$0.39          \\
 $ q       $                          & 0.22 $\pm$0.05   & 0.21 $\pm$0.05  &  0.21 $\pm$0.05          \\
 $ i\degr  $                          & 73.27$\pm$2      & 73.32$\pm$2     &  73.28$\pm$2           \\
 \hline
 \end{tabular}
\end{table} 
\onecolumn

\end{document}